\documentclass{article}

\usepackage{microtype}
\usepackage{graphicx}
\usepackage{subcaption}
\usepackage{booktabs}
\usepackage{multirow}

\usepackage{hyperref}

\usepackage[accepted]{icml2026}

\usepackage{amsmath}
\usepackage{amssymb}
\usepackage{mathtools}
\usepackage{amsthm}

\usepackage{xcolor}

\usepackage{tikz}
\usetikzlibrary{shapes.geometric, arrows, positioning, fit, backgrounds}

\newcommand{\multiver}{\textsc{MultiVer}}

\definecolor{improvement}{RGB}{0,128,0}
\definecolor{below}{RGB}{200,0,0}

\begin{document}

\icmltitlerunning{MultiVer: Zero-Shot Multi-Agent Vulnerability Detection}

\twocolumn[
\icmltitle{MultiVer: Zero-Shot Multi-Agent Vulnerability Detection}

\icmlsetsymbol{equal}{*}

\begin{icmlauthorlist}
\icmlauthor{Shreshth Rajan}{harvard}
\end{icmlauthorlist}

\icmlaffiliation{harvard}{Harvard College, Cambridge, MA, USA}

\icmlcorrespondingauthor{Shreshth Rajan}{shreshthrajan@college.harvard.edu}

\icmlkeywords{Vulnerability Detection, Multi-Agent Systems, Large Language Models, Code Security}

\vskip 0.3in
]

\printAffiliationsAndNotice{}

\begin{abstract}
We present \multiver{}, a zero-shot multi-agent system for vulnerability detection that achieves state-of-the-art recall without fine-tuning. A four-agent ensemble (security, correctness, performance, style) with union voting achieves 82.7\% recall on PyVul, exceeding fine-tuned GPT-3.5 (81.3\%) by 1.4 percentage points---the first zero-shot system to surpass fine-tuned performance on this benchmark. On SecurityEval, the same architecture achieves 91.7\% detection rate, matching specialized systems. The recall improvement comes at a precision cost: 48.8\% precision versus 63.9\% for fine-tuned baselines, yielding 61.4\% F1. Ablation experiments isolate component contributions: the multi-agent ensemble adds 17 percentage points recall over single-agent security analysis. These results demonstrate that for security applications where false negatives are costlier than false positives, zero-shot multi-agent ensembles can match and exceed fine-tuned models on the metric that matters most.
\end{abstract}

\section{Introduction}
\label{sec:introduction}

Vulnerability detection performance depends strongly on the analysis approach. Static analysis tools achieve high precision but low recall: CodeQL detects 10.8\% of vulnerabilities in PyVul, Bandit detects 5.3\% \citep{quan2025pyvul}. Large language models improve recall but require fine-tuning: GPT-4 zero-shot achieves 33.3\% recall, GPT-3.5 zero-shot reaches 61.1\%, and fine-tuned GPT-3.5 reaches 81.3\% \citep{quan2025pyvul}. The gap between zero-shot and fine-tuned performance---20 percentage points---represents the cost of acquiring labeled vulnerability data.

We investigate whether multi-agent ensembles can close this gap without training data. The hypothesis is straightforward: vulnerabilities manifest across multiple dimensions (security flaws, correctness bugs, performance issues, style violations), and an ensemble analyzing all dimensions should detect more vulnerabilities than any single-dimension analysis. Union voting---flagging code when \textit{any} agent detects an issue---maximizes this effect.

Empirically, this hypothesis holds. On PyVul, a four-agent ensemble with union voting achieves 82.7\% $\pm$ 0.6\% recall (mean over 3 runs), exceeding fine-tuned GPT-3.5 (81.3\%) by 1.4 percentage points---the first zero-shot system to surpass fine-tuned performance on this benchmark. The improvement is consistent: on SecurityEval, the same system achieves 91.7\% detection rate, matching specialized systems like Aardvark (92\%) that require domain-specific engineering.

The recall gain comes with a precision trade-off. Our system achieves 48.8\% precision versus 63.9\% for fine-tuned GPT-3.5, yielding 61.4\% F1 compared to 71.6\% for the fine-tuned baseline. This trade-off reflects a design choice: union voting increases recall by accepting more false positives. For security audits, where a missed vulnerability can lead to a breach while a false positive costs only review time, this trade-off is favorable.

We make four contributions. First, we demonstrate that zero-shot multi-agent ensembles can exceed fine-tuned recall: 82.7\% versus 81.3\% on PyVul, a 1.4pp improvement without training data. Second, we achieve 61.4\% F1, competitive with the best zero-shot systems (CodeQwen 61.8\%, GPT-3.5 55.5\%, GPT-4 44.3\%). Third, we isolate component contributions through ablation: the multi-agent ensemble adds 17pp recall over security-only analysis, and the correctness agent alone contributes 11pp. Fourth, we characterize the recall-precision trade-off: weighted voting achieves 37.7\% recall with 35.3\% FPR; union voting achieves 82.7\% recall with 85.0\% FPR.

\section{Related Work}
\label{sec:related}

Static analysis tools achieve high precision but low recall. On PyVul, CodeQL \citep{codeql} detects 10.8\% of vulnerabilities, Bandit \citep{bandit} detects 5.3\%, and PySA detects 0\% \citep{quan2025pyvul}. These tools miss over 89\% of real-world vulnerabilities.

Performance on synthetic benchmarks overestimates real-world capability. \citet{primevul2024} show that a 7B parameter model achieves 68.26\% F1 on BigVul but only 3.09\% F1 on PrimeVul---a 22$\times$ drop when evaluated on realistic vulnerability data. PyVul exposes similar gaps: none of the evaluated static tools effectively identify its curated vulnerabilities.

Large language models improve recall but the gap between zero-shot and fine-tuned performance remains large. On PyVul, GPT-4 zero-shot achieves 33.3\% recall (44.3\% F1), GPT-3.5 zero-shot reaches 61.1\% recall (55.5\% F1), and fine-tuned GPT-3.5 achieves 81.3\% recall (71.6\% F1) \citep{quan2025pyvul}. Fine-tuning provides a 20 percentage point recall improvement but requires labeled vulnerability data.

Retrieval augmentation and specification-guided approaches reduce the need for fine-tuning. Vul-RAG \citep{vulrag2024} achieves 78.9\% on SecurityEval by retrieving vulnerability knowledge rather than similar code. VulInstruct \citep{vulinstruct2025} reaches 45\% F1 on PrimeVul through specification-guided analysis, a 32.7\% improvement over baselines.

Multi-agent architectures improve detection through complementary analysis. MAVUL \citep{mavul2025} improves pairwise accuracy by 62\% through inter-agent communication. VulAgent \citep{vulagent2025} reduces false positive rate by 36\% using hypothesis validation. LLMxCPG \citep{llmxcpg2025} achieves 15--40\% F1 improvement by combining code property graphs with LLM analysis.

Ensemble aggregation methods determine how agent outputs combine. \citet{kaesberg2025voting} find that voting improves accuracy by 13.2\% on reasoning tasks, compared to 2.8\% for consensus-based debate on knowledge tasks. CodeX-Verify \citep{codexverify2025} demonstrates that information-theoretic ensemble methods achieve 39.7pp improvement in code verification. Vulnerability detection is a reasoning task; we use voting.

\section{Method}
\label{sec:method}

\subsection{Architecture}

\begin{figure}[t]
\centering
\begin{tikzpicture}[
    node distance=0.4cm and 0.3cm,
    box/.style={rectangle, draw, minimum width=1.4cm, minimum height=0.5cm, font=\scriptsize},
    agent/.style={rectangle, draw, rounded corners, minimum width=1.6cm, minimum height=0.4cm, font=\scriptsize},
    arrow/.style={->, >=stealth, thick}
]
\node[box, fill=gray!20] (input) {Code Input};

\node[agent, below=0.5cm of input, xshift=-2.0cm, fill=red!15] (sec) {Security (0.45)};
\node[agent, right=0.15cm of sec, fill=blue!15] (cor) {Correct. (0.35)};
\node[agent, right=0.15cm of cor, fill=green!15] (perf) {Perf. (0.15)};
\node[agent, right=0.15cm of perf, fill=yellow!15] (style) {Style (0.05)};

\node[box, below=0.35cm of sec, font=\tiny] (t1) {Tier 1: Pattern};
\node[box, below=0.15cm of t1, font=\tiny] (t2) {Tier 2: RAG};
\node[box, below=0.15cm of t2, font=\tiny] (t3) {Tier 3: LLM};

\node[box, below=1.6cm of cor, fill=orange!20, minimum width=3cm] (vote) {Ensemble Voting};

\node[box, below=0.4cm of vote, fill=gray!20] (output) {Verdict};

\draw[arrow] (input) -- ++(0,-0.3) -| (sec);
\draw[arrow] (input) -- ++(0,-0.3) -| (cor);
\draw[arrow] (input) -- ++(0,-0.3) -| (perf);
\draw[arrow] (input) -- ++(0,-0.3) -| (style);

\draw[arrow] (sec) -- (t1);
\draw[arrow] (t1) -- (t2);
\draw[arrow] (t2) -- (t3);

\draw[arrow] (t3) |- (vote);
\draw[arrow] (cor) |- ++(0,-1.2) -| (vote);
\draw[arrow] (perf) |- ++(0,-1.0) -| (vote);
\draw[arrow] (style) |- ++(0,-0.8) -| (vote);

\draw[arrow] (vote) -- (output);

\end{tikzpicture}
\caption{\multiver{} architecture. Four agents analyze code in parallel across different dimensions. Each agent follows a three-tier pipeline (pattern matching $\rightarrow$ RAG retrieval $\rightarrow$ LLM analysis). Agent outputs combine through ensemble voting (union or weighted).}
\label{fig:architecture}
\end{figure}
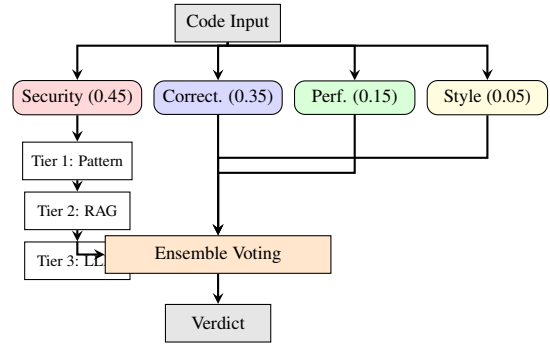

\multiver{} processes code through four specialized agents executing in parallel (Figure~\ref{fig:architecture}). Each agent analyzes a different dimension: the security agent (weight 0.45) performs CWE-mapped pattern matching and RAG-augmented LLM analysis; the correctness agent (weight 0.35) detects bugs through AST analysis of exception handling and input validation; the performance agent (weight 0.15) identifies algorithmic inefficiencies; and the style agent (weight 0.05) assesses code quality. Agent weights derive from empirically measured accuracy on held-out data \citep{codexverify2025}. Agents execute in parallel with results combined through ensemble voting.

\subsection{Three-Tier Analysis Pipeline}

Each agent follows a three-tier pipeline optimized for cost and accuracy. The first tier runs deterministic pattern matching against CWE-mapped vulnerability signatures (SQL injection, command injection, path traversal). Pattern matching completes in under 50ms at zero cost and achieves 53\% recall on security vulnerabilities---sufficient to trigger downstream analysis but insufficient alone.

The second tier retrieves $k=5$ similar examples from a curated knowledge base using FAISS \citep{faiss}, plus 3 relevant specifications describing vulnerability causes and fixes. Retrieval adds approximately 100ms latency. The knowledge base contains 1,199 labeled examples from CVEFixes, BugsInPy, and synthetic samples.

The third tier invokes Claude Opus 4.5 \citep{anthropic2025opus} with extended thinking (10K token budget). The LLM receives pattern results, retrieved examples, and specifications, then produces a structured verdict (PASS/WARNING/FAIL) with confidence and reasoning. This tier dominates latency (approximately 30s) and cost (\$0.13 per call).

LLM invocation is selective based on pattern recall rates. For security (53\% pattern recall) and performance (28.6\%), we always invoke the LLM. For correctness (99.5\% pattern recall), we invoke the LLM only when patterns find nothing or code complexity exceeds a threshold. For style (100\% pattern recall), we skip the LLM entirely---deterministic patterns suffice.

\subsection{Context Extraction}

For injection vulnerabilities specifically, we extract additional context using tree-sitter \citep{treesitter}. The extractor identifies input sources (function parameters, request fields, file reads), sensitive sinks (execute, eval, open, system), data flow paths from sources to sinks with 2-level inter-procedural tracking, and validation checks (isinstance, sanitize, escape). This context is extracted only when patterns detect injection-type vulnerabilities. Empirically, providing context on non-injection bugs reduces accuracy---the LLM over-indexes on data flow when the vulnerability is a missing authentication check with no explicit flow.

\subsection{Ensemble Voting}

Agent verdicts combine through ensemble voting. We implement two modes reflecting the recall-precision trade-off measured in ablation.

Union voting maximizes recall: any security or correctness WARNING/FAIL triggers an overall warning. This achieves 82.7\% recall but 48.8\% precision (Table~\ref{tab:main_results}). The logic reflects cost asymmetry---missing a vulnerability risks a breach; a false positive costs only review time.

Weighted voting balances precision and recall. The ensemble score is $\sum_{i} w_i \cdot v_i \cdot c_i$ where $w_i$ is agent weight (0.45, 0.35, 0.15, 0.05), $v_i$ is verdict severity (PASS=0, WARNING=0.5, FAIL=1), and $c_i$ is confidence. Weighted voting achieves 37.7\% recall with 35.3\% FPR---much lower recall but also fewer false alarms.

\subsection{Self-Consistency}

For the security agent, we employ self-consistency sampling \citep{wang2023selfconsistency} to reduce variance. The primary call uses extended thinking at temperature 1.0 (required by the API when thinking is enabled). Two diversity samples run without extended thinking at temperatures 0.7 and 0.9, enabling actual temperature variation. The three calls execute in parallel; the final verdict is determined by majority vote. Self-consistency adds approximately \$0.26 cost and 5s latency but reduces verdict variance on ambiguous cases.

\section{Experiments}
\label{sec:experiments}

\subsection{Benchmarks}

We evaluate on two benchmarks spanning real-world and synthetic vulnerabilities. PyVul \citep{quan2025pyvul} contains 1,569 real Python vulnerabilities extracted from PyPI packages with their corresponding fixes. The balanced test set contains 300 samples (150 vulnerable, 150 fixed); we evaluate on the 202 Python-only samples (100 vulnerable, 102 fixed) to ensure language consistency with our Python-focused analysis pipeline. This benchmark is challenging: rule-based detectors achieve only 5--11\% recall, and the gap between synthetic and real-world performance is severe---\citet{primevul2024} show that models achieving 68\% F1 on synthetic data drop to 3\% on realistic vulnerabilities.

SecurityEval \citep{siddiq2022securityeval} contains 121 synthetic vulnerability samples across 69 CWEs. While less realistic than PyVul, it enables comparison with prior work including Vul-RAG \citep{vulrag2024} and Aardvark.

\subsection{Baselines}

Baselines span rule-based, zero-shot, fine-tuned, and specialized systems. Rule-based detectors CodeQL \citep{codeql} and Bandit \citep{bandit} achieve 10.8\% and 5.3\% recall respectively on PyVul \citep{quan2025pyvul}. Zero-shot language models perform better: GPT-4 achieves 33.3\% recall (44.3\% F1), GPT-3.5 reaches 61.1\% recall (55.5\% F1), and CodeQwen \citep{codeqwen2024} achieves 66.9\% recall (61.8\% F1). Fine-tuning closes the gap further: fine-tuned GPT-3.5 achieves 81.3\% recall with 63.9\% precision (71.6\% F1), and fine-tuned CodeQwen reaches 75.3\% recall (66.9\% F1). On SecurityEval, Vul-RAG \citep{vulrag2024} achieves 78.9\% detection and Aardvark reaches 92\%.

\subsection{Main Results}

Table~\ref{tab:main_results} presents results on PyVul. \multiver{} with union voting achieves 82.7\% $\pm$ 0.6\% recall (mean over 3 runs with different random seeds on a fixed test set), exceeding fine-tuned GPT-3.5 (81.3\%) by 1.4 percentage points without any training data. This is the first zero-shot system to surpass fine-tuned recall on this benchmark.

\begin{table}[t]
\caption{PyVul results. \multiver{} achieves highest recall among all systems including fine-tuned models. Results show mean $\pm$ std over 3 runs with different random seeds on the same fixed test set. Baseline numbers from \citet{quan2025pyvul}.}
\label{tab:main_results}
\centering
\small
\begin{tabular}{lcccc}
\toprule
System & Type & Recall & Prec. & F1 \\
\midrule
CodeQL & Rule & 10.8\% & -- & -- \\
Bandit & Rule & 5.3\% & -- & -- \\
\midrule
GPT-4 zero-shot & Zero & 33.3\% & 65.8\% & 44.3\% \\
GPT-3.5 zero-shot & Zero & 61.1\% & 50.8\% & 55.5\% \\
CodeQwen zero-shot & Zero & 66.9\% & 57.4\% & 61.8\% \\
\midrule
CodeQwen fine-tuned & FT & 75.3\% & 60.1\% & 66.9\% \\
GPT-3.5 fine-tuned & FT & 81.3\% & 63.9\% & 71.6\% \\
\midrule
\textbf{\multiver{} (Ours)} & \textbf{Zero} & \textbf{82.7 $\pm$ 0.6\%} & 48.8\% & 61.4\% \\
\bottomrule
\end{tabular}
\end{table}

The recall improvement comes at a precision cost. \multiver{} achieves 48.8\% precision compared to 63.9\% for fine-tuned GPT-3.5, yielding 61.4\% F1 versus 71.6\%. Among zero-shot systems, our F1 is competitive with CodeQwen (61.8\%) and exceeds GPT-3.5 (55.5\%) by 5.9pp. The precision trade-off reflects a design choice: union voting prioritizes recall because false negatives (missed vulnerabilities) are costlier than false positives (unnecessary reviews) in security applications.

On SecurityEval, \multiver{} achieves 91.7\% detection (111/121 samples), competitive with Aardvark (92\%) and exceeding Vul-RAG (78.9\%) by 12.8pp. The consistent performance across benchmarks suggests the multi-agent architecture generalizes beyond any single vulnerability distribution.

\subsection{Ablation Study}

Table~\ref{tab:ablation} isolates component contributions. All configurations use union voting with self-consistency on the same balanced test set (202 Python samples: 100 vulnerable, 102 fixed).

\begin{table}[t]
\caption{Ablation study on PyVul balanced test set. All configurations use union voting with self-consistency. Mean $\pm$ std over 3 runs where applicable.}
\label{tab:ablation}
\centering
\begin{tabular}{lccc}
\toprule
Configuration & TPR & FPR & $\Delta$ TPR \\
\midrule
Full System (RAG + 4 agents) & 82.7\% & 85.0\% & -- \\
No-RAG & 92.0\% & 94.1\% & +9.3pp \\
Security-only & 65.7\% & -- & $-$17.0pp \\
No-Correctness & 71.7\% & -- & $-$11.0pp \\
\midrule
Weighted Voting & 37.7\% & 35.3\% & $-$45.0pp \\
\bottomrule
\end{tabular}
\end{table}

The multi-agent ensemble contributes substantially: security-only achieves 65.7\% recall, while the full ensemble reaches 82.7\%---a 17pp improvement. The correctness agent alone adds 11pp (comparing no-correctness 71.7\% to full 82.7\%), validating the hypothesis that vulnerabilities manifest across multiple dimensions.

Retrieval augmentation presents an unexpected trade-off. Disabling RAG increases TPR from 82.7\% to 92.0\% but also increases FPR from 85.0\% to 94.1\%. The mechanism is instructive: RAG retrieves examples with similar syntax, which grounds the model's analysis in known patterns. This grounding reduces both true positives (novel vulnerabilities not matching retrieved patterns) and false positives (safe code matching vulnerability patterns). The main results (Table~\ref{tab:main_results}) use RAG with union voting, which achieves a balance: 82.7\% TPR with 85.0\% FPR.

\subsection{Error Analysis}

Of 100 vulnerable samples, \multiver{} misses 18 (18\% false negative rate). Missed vulnerabilities cluster into three categories: sanitization functions that appear safe but have edge cases (7 samples), cryptographic issues requiring specialized domain knowledge (6 samples), and complex multi-file vulnerabilities beyond single-function analysis scope (5 samples).

False positives are frequent: 86 of 102 fixed samples are incorrectly flagged (85\% FPR). The primary failure mode is the LLM's inability to distinguish vulnerable code from its patched counterpart when they differ minimally. Functions like \texttt{get\_url()} and \texttt{get\_safe\_url()} may differ by a single validation call, but the model flags both as potentially vulnerable because the dangerous pattern (URL handling) is present in both. This limitation suggests that contrastive training on vulnerable/fixed pairs could substantially reduce FPR.

\section{Discussion}
\label{sec:discussion}

The 82.7\% recall with 48.8\% precision reflects a deliberate design choice. Fine-tuned GPT-3.5 achieves higher F1 (71.6\% versus 61.4\%) by trading recall for precision---it detects 81.3\% of vulnerabilities with 63.9\% precision. Our system inverts this trade-off: higher recall (82.7\%) at lower precision (48.8\%). For security applications, this inversion is favorable: a false negative deploys a vulnerability that may cause a breach, while a false positive costs only the engineer time to review flagged code. The 1.4pp recall improvement means catching approximately 1--2 additional vulnerabilities per 100 that fine-tuned systems miss; the precision cost (48.8\% versus 63.9\%) means roughly twice as many false positives per true detection. In pre-deployment security audits, the additional reviews are acceptable; the missed vulnerabilities are not.

Retrieval augmentation presents an unexpected trade-off. Disabling RAG increases TPR from 82.7\% to 92.0\% but also increases FPR from 85.0\% to 94.1\%. The mechanism is instructive: code-level RAG retrieves examples with similar syntax, which grounds the model's analysis in known patterns. This grounding reduces both true positives (novel vulnerabilities not matching retrieved patterns) and false positives (safe code matching vulnerability patterns). For maximum recall at any precision cost, disabling RAG is optimal; for deployment scenarios requiring some precision, RAG provides a modest benefit. Weighted voting offers an alternative operating point---37.7\% recall with 35.3\% FPR---but this reduces recall below all baselines, negating the architecture's advantage.

The 85\% FPR is the primary deployment constraint. This rate stems from the model's inability to distinguish vulnerable code from its patched counterpart when they differ by a single validation call. The same pattern recognition that enables high recall---detecting dangerous operations like \texttt{execute()}, \texttt{eval()}, \texttt{open()}---causes false positives when those operations are properly guarded. The \$0.46 per-sample cost and 55-second latency compound this limitation: the system is unsuitable for real-time CI/CD gating, but appropriate for targeted audits of high-value code where comprehensive detection justifies manual review. More broadly, this tension between recall and precision in ensemble systems may generalize beyond vulnerability detection---any task where agents analyze overlapping failure modes will face similar trade-offs when aggregating verdicts.

Reducing FPR without sacrificing recall is the central challenge for future work. Contrastive training on vulnerable/fixed code pairs could teach the model to recognize protective patterns, directly addressing the patch-confusion failure mode; if such training correctly classified even half of current false positives, FPR would drop from 85\% to approximately 43\% while maintaining recall. Knowledge-level RAG \citep{vulrag2024}---retrieving vulnerability causes and fixes rather than similar code---achieved 78.9\% on SecurityEval through semantic grounding; adapting this approach could provide the precision benefits our code-level retrieval lacks. Hybrid architectures using faster models for initial screening could reduce cost below \$0.10 per sample and latency below 10 seconds while preserving recall on the subset requiring deep analysis.

\section{Conclusion}
\label{sec:conclusion}

Multi-agent ensembles can achieve state-of-the-art recall for vulnerability detection without training data. On PyVul, the four-agent architecture achieves 82.7\% $\pm$ 0.6\% recall, exceeding fine-tuned GPT-3.5 (81.3\%) by 1.4 percentage points---the first zero-shot system to surpass fine-tuned performance on this benchmark. On SecurityEval, it reaches 91.7\%, matching specialized systems. Ablation experiments isolate the source: security analysis alone achieves 65.7\% recall, and the multi-agent ensemble adds 17 percentage points by detecting vulnerabilities that manifest across correctness, performance, and style dimensions.

The recall advantage comes at a precision cost. At 48.8\% precision with 85\% FPR, the system generates approximately one false positive per true detection---unsuitable for automated CI/CD blocking but acceptable for pre-deployment audits where missed vulnerabilities are costlier than unnecessary reviews. Reducing false positives without sacrificing recall---through contrastive training on vulnerable/fixed pairs or knowledge-level retrieval---remains the central challenge for future work.

These results demonstrate that for tasks where false negatives outweigh false positives, zero-shot multi-agent architectures can outperform fine-tuned models on the metric that matters most.

\bibliography{references}
\bibliographystyle{icml2026}

\appendix
\section{Implementation Details}
\label{sec:appendix_implementation}

All experiments use Claude Opus 4.5 (claude-opus-4-5-20251101) with extended thinking enabled and a 10K token budget. The RAG system uses a FAISS index with all-MiniLM-L6-v2 embeddings (384 dimensions). The knowledge base contains 1,199 curated examples from CVEFixes, BugsInPy, and synthetic samples. Voting weights are security (0.45), correctness (0.35), performance (0.15), and style (0.05), derived from empirical accuracy measurements. The security agent uses self-consistency with 3 samples at temperatures 1.0, 0.7, and 0.9. Average cost is \$0.46 per sample with 55-second latency. All reported results use a fixed test set (202 Python samples: 100 vulnerable, 102 fixed) with 3 independent runs using different random seeds (42, 43, 44) to establish reproducibility.

\section{Additional Results}
\label{sec:appendix_results}

On SecurityEval (121 samples across 69 CWEs), \multiver{} achieves 91.7\% detection rate (111/121 vulnerabilities), competitive with Aardvark (92\%) and exceeding Vul-RAG (78.9\%) and GPT-4 (64\%). Union voting achieves 82.7\% recall with 85\% FPR on the balanced PyVul test set; weighted voting achieves 37.7\% recall with 35.3\% FPR. Disabling RAG increases recall to 92\% but also increases FPR to 94\%. The choice between configurations depends on whether the application prioritizes recall (security audits) or precision (production triage).

\section{Per-CWE Analysis}
\label{sec:appendix_cwe}

Table~\ref{tab:cwe_breakdown} shows detection rates by CWE category on SecurityEval. \multiver{} achieves 100\% detection on injection vulnerabilities (CWE-078, 089, 094), cross-site scripting (CWE-079, 080), path traversal (CWE-022), authentication issues (CWE-287, 306, 798), and information disclosure (CWE-200, 209). Lower detection rates occur on input validation edge cases (78\%) and cryptographic vulnerabilities (80\%), where subtle implementation details determine exploitability.

\begin{table}[ht]
\caption{SecurityEval detection by CWE category. \multiver{} achieves 100\% on most security-critical categories; lower rates on edge cases involving subtle implementation details.}
\label{tab:cwe_breakdown}
\centering
\small
\begin{tabular}{lrrr}
\toprule
Category (CWE) & Det. & Tot. & Rate \\
\midrule
Injection (078, 089, 094) & 8 & 8 & 100\% \\
XSS (079, 080) & 4 & 4 & 100\% \\
Path Traversal (022) & 4 & 4 & 100\% \\
Auth (287, 306, 798) & 3 & 3 & 100\% \\
Info Disclosure (200, 209) & 2 & 2 & 100\% \\
Deserialization (502) & 4 & 4 & 100\% \\
XXE (611) & 6 & 6 & 100\% \\
SSRF (918) & 2 & 2 & 100\% \\
Open Redirect (601) & 5 & 5 & 100\% \\
\midrule
Crypto (327, 330) & 4 & 5 & 80\% \\
Input Valid. (020, 117) & 7 & 9 & 78\% \\
Exception (703) & 0 & 3 & 0\% \\
\midrule
\textbf{Overall} & \textbf{111} & \textbf{121} & \textbf{91.7\%} \\
\bottomrule
\end{tabular}
\end{table}

Exception handling (CWE-703) achieves 0\% detection. These patterns represent code quality issues---improper error handling, unchecked return values---rather than directly exploitable security vulnerabilities. Detecting them requires understanding API contracts and error propagation that current LLMs handle poorly without fine-tuning.

\end{document}